\documentclass[12pt,a4paper]{article}
\usepackage{epsf,feynmf}
\newcommand{\cd}{\makebox[0.08cm]{$\cdot$}} \title{\bf {The charge and
mass perturbative renormalization in explicitly covariant LFD}}
\author{J.-J. Dugne$^a$\thanks{e-mail: dugne@clermont.in2p3.fr},
\and V.A. Karmanov$^b$\thanks{e-mail: karmanov@sci.lebedev.ru}, \and
J.-F. Mathiot$^a$\thanks{e-mail: mathiot@in2p3.fr} \\ \\
{\small \em $^a)$ Laboratoire de Physique Corpusculaire,
Universit\'e Blaise Pascal,} \\ {\small \em CNRS/IN2P3, 24 avenue des
Landais, F-63177 Aubi\`ere Cedex, France} \\ {\small \em
$^b)$ Lebedev Physical Institute, Leninsky Prospekt 53, 117924 Moscow,
Russia} }

\date{}

\begin{document}
\begin{fmffile}{graphesJF}

\maketitle
\bibliographystyle{unsrt}

\abstract{In several preceding studies, the explicitly covariant
formulation of light front dynamics was developed and applied to many
observables.  In the present study we show how in this approach the
renormalization procedure for the first radiative correction can be
carried out in a standard way, after separating out the contributions
depending on the orientation of the light-front plane.  We calculate
the renormalized QED vertices $\gamma e^-\to e^-$ and $e^+e^-\to
\gamma $ and the electron self-energy and recover, in a
straightforward way, the well known analytical results obtained in the
>Feynman approach.}

\vspace{2cm}
\noindent  PACS numbers: 11.10.Gh, 11.25.Db, 11.30.Cp\\
\noindent  Keywords: Renormalization, Light Front Dynamics\\
\noindent  PCCF RI 00-13\\

\newpage

%%%%%%%%%%%%%%%%%%%%%%%%%%%%%%%%%%%%%%%%%%%%%%%%%%%%%%%%%%%%%%%%%
\section{Introduction}

Light Front Dynamics is a field theoretical approach which has been
successfully applied to relativistic composite systems.  The
two forms of this scheme are: the standard Light Front Dynamics (LFD)
\cite{bpp} and explicitly Covariant Light Front
Dynamics (CLFD) \cite{cdkm}.  While the standard LFD deals with the
state vector defined on the plane $t+z=0$, this plane is defined in
CLFD by the invariant equation $\omega\cd x=0$, where $\omega$ is a
four-vector with $\omega^2=0$. The particular choice of the
four-vector $\omega=(1,0,0,-1)$ turns CLFD into standard LFD.

In this article, we apply CLFD to the calculation, in first order
perturbation theory, of the QED $\gamma e^-\to e^-$ and $e^+e^-\to
\gamma $ vertices and of the fermion self-energy.  We will illustrate
in details the calculational techniques of CLFD in order to point out
its differences and similarities with respect to standard LFD and to
the Feynman techniques.

The main difference with calculations in the Feynman approach lies
in the fact that in CLFD, like in ordinary LFD, all the four-momenta,
even in the intermediate states, are on their mass shells, whereas the
amplitudes may be off-energy shells.  Moreover, the amplitude may
depend, in a well defined manner, on the orientation of the
light-front plane, i.e., on the four-vector $\omega$.  This dependence
is an unphysical one for physical amplitudes.  Thus explicit covariance
allows to  disentangle clearly the physical amplitudes from unphysical
ones.  In this respect, the CLFD calculations differ from the
standard LFD ones.

The study of the perturbative renormalization
has already been done in standard LFD
\cite{brs73}-\cite{glaz92}, resulting in the non-locality of the
necessary counterterms. In our covariant approach this nonlocality
manifests itself only in the terms depending on
the orientation of the light front plane. These terms can be
explicitly
removed from the physical amplitude.  We will show that
after their separation, the renormalization of the $\omega$-independent
part of the amplitude is carried out in a very simple way, like in the Feynman
approach, and does not require any  non-local counterterms.
  Then we find that the on-energy shell
  electromagnetic vertex
in CLFD coincides with the on-mass shell Feynman vertex.
The same is true for the electron self-energy.

In sect.2,
we start with the calculation of the electron electromagnetic vertex in
  CLFD: the anomalous magnetic moment of the electron, and the renormalized
  electron charge.  In sect.\ref{epem}
we apply our formalism to the vertex $e^+e^-\to \gamma$ for the threshold
value of the photon momentum $Q^2=4m^2$, and discuss the physical infrared
singularity. The renormalized electron mass operator
is calculated in sect.\ref{sen}.
Sect.\ref{concl} contains our concluding remarks.
Some technical details are given in the appendices \ref{ap1} and \ref{ap2}.

%%%%%%%%%%%%%%%%%%%%%%%%%%%%%%%%%%%%%%%%%%%%%%%%%%%%%%%%%%%%%%%%%
\section{The electron electromagnetic vertex}

\subsection {The anomalous magnetic moment of the electron}\label{anomal}
The anomalous magnetic moment of the electron is a simple example of a
higher order process in QED. Its calculation gives a finite result
and does not require renormalization. It
allows us also to show how to disentangle  $\omega$-dependent terms
in our formalism.

The spin 1/2 electromagnetic vertex in CLFD has the general form:
\begin{equation}\label{eq12}
J_{\rho}(q)= \bar{u}(p')\Gamma_{\rho}u(p)\ ,
\end{equation}
where $q=p'-p$. We shall denote $Q^2=-q^2$. Due to the explicit
covariance of our approach, the vertex operator
$\Gamma_{\rho}$, according to ref.\cite{km96}, can be decomposed into:
\begin{equation}\label{eq2b}
\Gamma_{\rho}=F_1\gamma_{\rho}
+\frac{iF_2}{2m}\sigma_{\rho\nu}q^{\nu}
+B_1\left(\frac{\hat{\omega}}{\omega\cd p} -\frac{1}{(1+\eta)m}
\right)P_{\rho}
+B_2\frac{m}{\omega\cd p}\omega_{\rho}
+B_3\frac{m^2}{(\omega\cd p)^2}\hat{\omega}\omega_{\rho}\ ,
\end{equation}
where $\sigma^{\rho\nu}=i(\gamma^{\rho}\gamma^{\nu}
-\gamma^{\nu}\gamma^{\rho})/2$, $\hat{\omega}=\omega_\mu\gamma^\mu$,
$\eta=Q^2/(4m^2)$ and $m$ is the electron mass.  The electromagnetic
vertex (\ref{eq12}) is gauge invariant since $J_{\rho}q^{\rho}=0$
(with the condition $\omega \cd q =0$).  The possible
non-gauge-invariant terms are forbidden by $T$-invariance.  The
anomalous magnetic moment is the value of $F_2(Q^2)$ for $Q^2=0$.

The physical form factors $F_1$ and $F_2$  can easily be extracted from
the vertex function
$\Gamma_{\rho}$. To this end,  we multiply $J_{\rho}$ by
$ [\bar{u}^{\sigma'}(p')\gamma^{\rho}u^{\sigma}(p)]^*$,
$[\bar{u}^{\sigma'}(p')i\sigma^{\rho\nu}q_{\nu}/(2m)
u^{\sigma}(p)]^*$, etc. and sum over polarizations. After taking the
trace, we
obtain the following quantities:
\begin{eqnarray}\label{eq14n}
&&c_1=Tr[O_{\rho}\gamma^{\rho}]\ ,
\quad
c_2=Tr[O_{\rho}i\sigma^{\rho\nu}q_{\nu}]/(2m)\ ,
\quad
c_3=Tr[O_{\rho}(\hat{\omega}/\omega\cd p -1/(1+\eta)m)]P^{\rho}\ ,
\nonumber\\
&&c_4=Tr[O_{\rho}]\omega^{\rho}m/\omega\cd p\ ,
\quad
c_5=Tr[O_{\rho}\hat{\omega}]\omega^{\rho}m^2/(\omega\cd p)^2\ ,
\end{eqnarray}
where
\begin{equation}\label{eq14p}
O_{\rho}=(\hat{p}'+m)\Gamma_{\rho}(\hat{p}+m)/(4m^2)\ .
\end{equation}
With the decomposition (\ref{eq2b}) of $\Gamma_{\rho}$, we get a linear
system of five equations for $F_1,F_2,B_{1-3}$ with the inhomogeneous
part determined by $c_{1-5}$.  Solving this system relative to $F_2$,
we find:
\begin{equation}\label{eq19}
F_2=\frac{\displaystyle{1}}{\displaystyle{4\eta
(1+\eta)^2}}[(c_3+4c_4-2c_1)(1+\eta) +2(c_1+c_2)-2(c_5+c_4)
(1+\eta)^2]\ .
\label{eq20}
\end{equation}
In spite of $\eta$ in the denominator in eq.(\ref{eq20}),
there is no singularity at $Q^2=0$.

In the usual formulation of LFD on the plane $t+z=0$,  the  form
factors of spin 1/2 systems are found from  the plus-component of the current,
i.e., in our notation, from the
contraction of ${J}_{\rho}$ in eq.(\ref{eq12}), with $\omega_{\rho}$.
This contraction gets rid of the contributions of $B_{2,3}$, but
not the term proportional to $B_1$.  The form factors $F_1'$ and
$F_2'$ inferred in this way are thus given by:
\begin{eqnarray}
\label{eq22n} J \cd \omega&=&\bar{u}'[F_1\hat{\omega}
+\frac{iF_2}{2m}\sigma_{\rho\nu}\omega^{\rho}q^{\nu}
+2B_1(\hat{\omega}-\frac{\omega\cd p}{(1+\eta)m})]u \nonumber\\
&\equiv&\bar{u}'[ F_1'\gamma_{\rho}
+\frac{iF_2'}{2m}\sigma_{\rho\nu}q^{\nu}]u\ \omega^{\rho}\ ,
\end{eqnarray}
where
\begin{equation}\label{eq23a}
F_1'=F_1+\frac{2\eta B_1}{1+\eta}\ ,\quad
F_2'=F_2+\frac{2}{1+\eta}B_1\ .
\end{equation}

The  $B_1$ expression can be found from the above mentioned system of
equations, leading to:
\begin{equation}\label{eq24a}
B_1=-\frac{1}{8\eta (1+\eta)}[(c_3+4c_4-2c_1)(1+\eta)+2(c_1+c_2)-4c_5
(1+\eta)^2]\ .
\end{equation}
Substituting $F_2$ from (\ref{eq19}) and $B_1$ from (\ref{eq24a}) into
eq. (\ref{eq23a}) for $F'_2$, we get:
\begin{equation}\label{eq25n}
   F_2'=(c_5-c_4)/(2\eta)\ .
\end{equation}
Of course, in a given order of perturbation theory, both methods for 
calculating
the form
factors, by eqs.(\ref{eq19}) and (\ref{eq25n}), should give the same result.
This means  we should find $B_1=0$. We shall see below that it is
indeed the case.

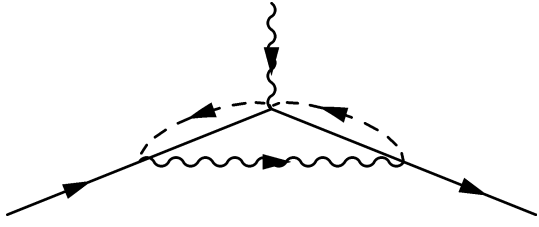
\begin{figure}[hbtp]
\vspace*{2cm}
\hspace*{4 cm}
\begin{fmfgraph*}(100,40)
\fmfipair{aa,bb,cc,dd,ee,ff}
\fmfiequ{dd}{sw}
\fmfiequ{aa}{dd+(.5w,.5h)}
\fmfiequ{cc}{aa+(.5w,.5h)}
\fmfiequ{ff}{cc+(0,1.h)}
\fmfiequ{bb}{cc+(.5w,-.5h)}
\fmfiequ{ee}{bb+(.5w,-.5h)}
\fmfi{photon,lab=$q$,lab.sid=left}{ff--cc}
\fmfi{photon,lab=$k$,lab.sid=right}{aa--bb}
\fmfi{phantom_arrow}{ff--cc}
\fmfi{phantom_arrow}{aa--bb}
\fmfi{photon,lab=$k$,lab.sid=right}{aa--bb}
\fmfi{fermion,lab=$p$,lab.sid=left}{dd--aa}
\fmfi{plain}{aa--cc}
\fmfi{plain}{cc--bb}
\fmfi{fermion,lab=$p$,lab.sid=left}{bb--ee}
\fmfi{dashes_arrow,lab=$\omega\tau_{1}$,lab.sid=right}{bb{up} .. tension
3 .. {down}cc}
\fmfi{dashes_arrow,lab=$\omega\tau_{2}$,lab.sid=right}{cc{up} .. 
tension 3 .. {down}aa}
\end{fmfgraph*}
\vspace*{1cm}
\begin{center}
\caption{\label{f2} Electromagnetic vertex of the electron. The dashed
line represents the spurion line, as explained in details in ref.\cite{cdkm}}
\end{center}
\end{figure}

Let us first calculate the form factor $F'_2$, for $Q^2=0$.
The amplitude corresponding to the graph of fig. \ref{f2}
is given by the rules  of the
graph techniques \cite{cdkm} and  has the form:
\begin{eqnarray}\label{eq1n}
\bar{u}(p)\Gamma^{\rho}u(p)&=&e^2\int\bar{u}(p)
\gamma_{\mu}(\hat{p}-\hat{k}+m)
\theta(\omega\cd(p- k))\delta((p+\omega\tau_1-k)^2-m^2)
\frac{d\tau_1}{\tau_1-i0}
\nonumber\\
&\times&  \gamma^{\rho}(\hat{p}-\hat{k}+m)
\theta(\omega\cd (p-k))\delta((p+\omega\tau_2-k)^2-m^2)
\frac{d\tau_2}{\tau_2-i0}\gamma_{\nu}u(p)
\nonumber\\
&\times&(-g^{\mu\nu})\theta(\omega\cd k)\delta(k^2-\mu^2)
\frac{d^4k}{(2\pi)^3}\ .
\end{eqnarray}
We use the Feynman gauge for the photon propagator.
The factor
$\hat{p}-\hat{k}=\hat{k}_1-\hat{\omega}\tau_1=\hat{k}_2-\hat{\omega}\tau_2$
includes the contact terms $-\hat{\omega}\tau_1$ and
$-\hat{\omega}\tau_2$,
as explained in ref.\cite{cdkm}. For the regu\-larization of subsequent
calculations,
we introduced in (\ref{eq1n}) the photon mass $\mu$, although it is not
necessary
in the present subsection.

Integrating over $\tau_{1},\tau_{2}$ and $k_0$, we get:
\begin{equation}\label{eq1a}
\bar{u}(p)\Gamma^{\rho}u(p)= e^2\int
\frac{\bar{u}(p)G^\rho u(p)}
{(s-m^2)^2(1-x)^2}
\frac{d^3k}{2\varepsilon_k(2\pi)^3}\ ,
\end{equation}
where
$s=(k+k_1)^2=(k+k_2)^2$, $x=\omega\cd k/\omega\cd p$ and  we note:
\begin{equation}\label{gam}
G^\rho=-\gamma_{\mu}(\hat{p}-\hat{k}+m)
\gamma^{\rho}(\hat{p}-\hat{k}+m)\gamma^{\mu}.
\end{equation}

The integrands for the scalar functions $c_{1-5}$ are represented
in terms of the
scalar products between the four-momenta $p$, $k$ and $\omega$. The scalar
product $p\cd k$  is given by:
$$
p\cd k =\mu^2/2 + (1-x)(s-m^2)/2\ ,
$$
whereas the scalar products $\omega\cd k$ and $\omega\cd p$ always appear
in the ratio $x$, with $0\le x \le 1$.

It is convenient to introduce the variable $R=k-xp$.
As usual (see, e.g., ref.\cite{cdkm}),
we represent the spatial part of
$R$ as $\vec{R}=\vec{R}_{\|}
+\vec{R_{\perp}}$, where $\vec{R}_{\|}$ is parallel to $\vec{\omega}$  and
$\vec{R}_{\perp}$
is orthogonal to $\vec{\omega}$.  Since,  by definition  of  $R$,
$R\cd\omega=R_0\omega_0-\vec{R}_{\|}\cd\vec{\omega}=0$,
it follows that $R_0=|\vec{R}_{\|}|$, and, hence,  $\vec{R}^2_{\perp} =-R^2$ is
invariant. In terms of
$\vec{R}^2_{\perp}$ and $x$, the variable $s$ writes:
\begin{equation}\label{s}
s=\frac{\vec{R}^2_{\perp}+\mu^2}{x}+ \frac{\vec{R}^2_{\perp}+m^2}{1-x}.
\end{equation}
and the integration volume is transformed as:
$d^3k/\varepsilon_k=d^2R_{\perp} dx/x$\ .

Substituting these expressions  into (\ref{eq1a}), we find (for $\mu=0$):
\begin{equation}\label{eq1b}
\bar{u}(p)\Gamma^\rho u(p)=\frac{\alpha}{4\pi^2}\int
\bar{u}(p)G^\rho u(p)
\frac{xdxd^2 R_{\perp}}{(\vec{R}^2_{\perp}+xm^2)^2}\ ,
\end{equation}
where we denote $\alpha=e^2/4\pi$.
To calculate $F'_2$ by eq.(\ref{eq25n}) (for $\eta\to 0$),
we substitute $\Gamma_\rho$ from (\ref{eq1b})
into (\ref{eq14p}) (for $p'=p$) and then into expressions
(\ref{eq14n}) for $c_4$ and $c_5$. Calculating  the traces, we get:
\begin{equation}\label{eq1c}
F'_2(0)=\frac{\alpha}{4\pi^2}\int 4m^2 x(1-x)
\frac{xdxd^2 R_{\perp}}{(\vec{R}^2_{\perp}+xm^2)^2}\ .
\end{equation}
We thus obtain the well known
result for the anomalous magnetic moment of the electron:
\begin{equation}\label{eq1d}
F'_2(0)=\frac{\alpha}{2\pi}\ .
\end{equation}

Now consider the form factor $F_2$
calculated after separation of the $\omega$-dependent terms. According to
(\ref{eq23a}), it is related to $F'_2$ by $F_2(0)=F'_2(0)-2B_1(0)$.
 From eq. (\ref{eq24a}), for $Q^2=0$,
we find the following expression for $B_1$:
\begin{equation}\label{eqb1}
B_1(0)=\frac{\alpha}{2\pi}\int 
\frac{[m^2(2-x)x^2-2R^2_{\perp}(1-x)-\mu^2(2-x)]}
{[R^2_{\perp}+m^2x^2+\mu^2(1-x)]^2}R_{\perp}dR_{\perp}dx\ ,
\end{equation}
which is logarithmically divergent. We regularize it using
the Pauli-Villars method, i.e., the photon propagator is replaced by:
\begin{equation}\label{eq7a}
\frac{1}{k^2-\mu^2}\rightarrow
\frac{1}{k^2-\mu^2}-\frac{1}{k^2-\Lambda^2}\ .
\end{equation}
In the absence of infrared singularity we can put in (\ref{eq7a})
$\mu=0$. Hence, the regularized expression for $B_1$ reads:
$$
B_1^{reg}=B_1(\mu=0)-B_1(\mu=\Lambda)\ .
$$
Integrating it over $R_{\perp}$, we get:
\begin{equation}\label{eqb3}
B_1^{reg}=-\frac{\alpha}{4\pi}\int_0^1 dx \frac{d}{dx}
\left[x(2-x)
\log\left(\frac{\Lambda^2 (1-x)+m^2 x^2}{m^2 x^2}\right)\right].
\end{equation}
After integration over $x$ we get $B_1^{reg}=0$ for any value of $\Lambda$.
This clearly shows that both methods to calculate the anomalous magnetic moment
of the electron give the same result.

%%%%%%%%%%%%%%%%%%%%%%%%%%%%%%%%%%%%%%%%%%%%%%%%%%%%%%%
\subsection{The renormalized electron charge}\label{renorm}

In order to calculate the radiative correction to
the form factor $F_1$, we have to renormalize the
charge. The renormalization means that the Lagrangian contains a counter
term of the form:
$$
Z_{1}\bar{\psi}\gamma^{\rho}\psi A_{\rho}\ ,
$$
hence, the amplitude $J^{\rho}$ is replaced by
\begin{equation}\label{eq21a}
J^{\rho}\rightarrow J^{\rho}_{ren}=J^{\rho}-J^{\rho}_0\ ,
\end{equation}
where
$$
J^{\rho}_0 = Z_{1}\bar{u}(p')\gamma^{\rho}u(p)\ .
$$
The renormalization procedure is described in many textbooks, see for
example \cite{IZ,MS,ll}.
In order to find $Z_{1}$, one must calculate the amplitude
$\bar{u}(p)\Gamma^{\rho}u(p)$ from the diagram of fig. \ref{f2}.
The value of $Z_{1}$ is in fact just the form factor $F_1(0)$ 
determined by this
diagram. For $p=p'$ the general decomposition (\ref{eq12}) turns into
\begin{equation}\label{eq23}
\bar{u}(p)\Gamma^{\rho}u(p)=Z_{1}\bar{u}(p)\gamma^{\rho}u(p)
+Z'\frac{\omega^{\rho}m}{\omega\cd p}\bar{u}(p)u(p)\ ,
\end{equation}
where $Z_{1}=F_1(0)$ and $Z'=B_2(0)+B_3(0)$.

  From (\ref{eq23}) the constant $Z_{1}$ is given by:
\begin{equation}
Z_{1}=\frac{1}{4\omega\cd
p}Tr\left[\omega_{\rho}\Gamma^{\rho}(\hat{p}+m)\right]\ .
\label{eq24}
\end{equation}
The vertex $\Gamma^\rho$ is determined by eq.(\ref{eq1n})
and is reduced to  (\ref{eq1a}). For  regularization purposes,
we should now keep the photon mass $\mu$ finite. From eq.(\ref{eq24}) we find:
\begin{equation}\label{eq29}
Z_{1}=\frac{\alpha}{(2\pi)^3}\int d^2R_{\perp}\int_0^1
\frac{\left[\vec{R}_{\perp}^2 + m^2(-2 + 2x + x^2)\right]x}
{\left[\vec{R}_{\perp}^2+m^2x^2+\mu^2(1-x)\right]^2}dx\ .
\end{equation}
The subsequent calculation is similar to the calculation
given in appendix \ref{ap1}.
That is, we calculate $Z_{1}(\mu,L)$ for a fixed upper limit $L$ of
the variable $R_{\perp}$ in the integral (\ref{eq29}),
take the difference $Z_{1}(\mu,L)-Z_{1}(\Lambda,L)$, take the limit
$L\to\infty$ and then calculate the limits $\mu\to 0$ and
$\Lambda\to\infty$. We then obtain:
\begin{equation}\label{eq30a}
Z_{1}(\mu\to 0,\Lambda\to \infty)=\frac{9\alpha}{8\pi}
+\frac{\alpha}{2\pi}\log\left(\frac{\mu^2}{m^2}\right)
+\frac{\alpha}{4\pi}\log\left(\frac{\Lambda^2}{m^2}\right)\ .
\end{equation}
This expression exactly coincides with the
expression found in the Feynman formalism \cite{correc}.
We emphasize that this result for $Z_{1}$ is obtained for the physical part
of the full vertex (\ref{eq23}), after separating out the unphysical 
term proportional
to $Z'\omega_{\rho}$. The latter term can be disregarded, there is no need
to calculate it.

%%%%%%%%%%%%%%%%%%%%%%%%%%%%%%%%%%%%%%%%%%%%%%%%%%%%%%%%%%%%%%%%%%
\section{Application to the vertex  $e^+e^- \rightarrow \gamma$}\label{epem}
As a direct application of the preceding calculation, let us now
consider the leptonic decay width of the positronium. In the
Weisskopf-Van Royen limit, the decay width is proportional to the
elementary vertex $  e^+e^-\rightarrow \gamma$, where the $e^+e^-$ pair
originates from the positronium wave function with zero relative
momentum, i.e. with $p_{e^+}=p_{e^-}=p$.

%%%%%%%%%%%%%%%%%%%%%%%%%%%%%%%%%%%%%%%%%%%%%%%%%%%%%%%%%%%%%%%%%%%%%%%%%%%%%%
\subsection{On-energy-shell spin structure}\label{spst}
The on-shell amplitude for the process $e^+e^- \rightarrow
\gamma $ depends on the four-vectors $p$ and $\omega$.  Its general
structure thus reads:
\begin{equation}\label{eq2}
\bar{u}(p)M^{\rho}v(p)=A\bar{u}(p)\gamma^{\rho}v(p)
+B \frac{p^{\rho}}{\omega\cd p}\bar{u}(p)\hat{\omega}v(p)
+C\frac{\omega^{\rho}m^2}{(\omega\cd p)^2}\bar{u}(p)\hat{\omega}v(p)\ .
\end{equation}
Here $v(p)$ is the positron spinor. The constant $A$ in (\ref{eq2}) is the
value of the form factor $F_1(Q^2)$ at $Q^2=4m^2$.
One can also construct the structure
$\sigma^{\rho\beta}\omega_{\beta}/\omega\cd p$,
but it is not independent, since:
$$
\frac{im}{\omega\cd p}\bar{u}(p)\sigma^{\rho\beta}\omega_{\beta}v(p)
=\bar{u}(p)\gamma^{\rho}v(p)-\frac{p^{\rho}}{\omega\cd p}
\bar{u}(p)\hat{\omega}v(p)\ .
$$

Multiplying (\ref{eq2}) on the left  by $u(p)$ and on the right by $\bar{v}(p)$
and
summarizing over polarizations, we get the factors $\sum_\lambda u(p)\bar{u}(p)
=(\hat{p}+m)$, $\sum_\lambda v(p)\bar{v}(p)=(\hat{p}-m)$.
We introduce therefore the quantity:
\begin{equation}\label{eq3}
\tilde{M}^{\rho}=(\hat{p}+m)M^{\rho}(\hat{p}-m)\ ,
\end{equation}
and calculate the following traces:
\begin{eqnarray}\label{eq4}
T_1&\equiv&\frac{1}{16m^2}Tr\left[\tilde{M}^{\rho}\gamma_{\rho}\right]
=(-3A+C)/2\ ,
\nonumber\\
T_2&\equiv&\frac{1}{16m^2}Tr\left[\tilde{M}^{\rho}\hat{\omega}\right]
\frac{p_{\rho}}{\omega\cd p}=(B+C)/2\ ,
\nonumber\\
T_3&\equiv&\frac{1}{16m^2}Tr\left[\tilde{M}^{\rho}\hat{\omega}\right]
\frac{\omega_{\rho}m^2}{(\omega\cd p)^2}=(A+B)/2\ .
\end{eqnarray}
So we can  find out the coefficients which determine the amplitude (\ref{eq2}):
\begin{eqnarray}\label{eq5}
A&=&T_2-T_1-T_3\ ,
\nonumber\\
B&=&T_1-T_2+3T_3\ ,
\nonumber\\
C&=&-T_1+3T_2-3T_3\ .
\end{eqnarray}

%%%%%%%%%%%%%%%%%%%%%%%%%%%%%%%%%%%%%%%%%%%%%%%%%%%%%%%%
\subsection{The physical amplitude}\label{ampl}
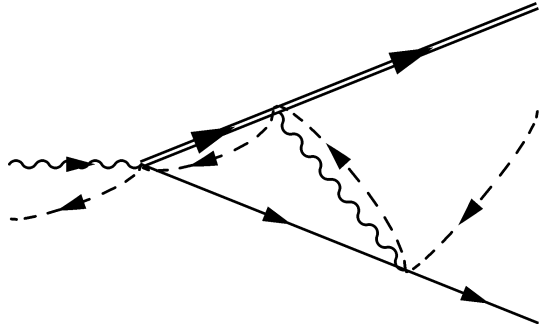
\begin{figure}[hbtp]
\vspace*{4cm}
\hspace*{4 cm}
\begin{fmfgraph*}(100,40)
\fmfipair{aa,bb,cc,dd,ee,ff,gg,ii}
\fmfiequ{aa}{0.5[sw,nw]}
\fmfiequ{bb}{aa+(.5w,0)}
\fmfiequ{cc}{bb+(.5w,.5h)}
\fmfiequ{dd}{bb+(w,-h)}
\fmfiequ{ee}{cc+(w,h)}
\fmfiequ{gg}{dd+(.5w,-.5h)}
\fmfiequ{ff}{ee+(0,-h)}
\fmfiequ{ii}{aa+(0,-.5h)}
\fmfi{photon}{cc--dd}
\fmfi{photon,lab=$q$,lab.sid=left}{aa--bb}
\fmfi{phantom_arrow}{aa--bb}
\fmfi{heavy,lab=$k_{1}$,lab.sid=left}{bb--cc}
\fmfi{heavy,lab=$p'$,lab.sid=left}{cc--ee}
\fmfi{fermion,lab=$k$,lab.sid=right}{bb--dd}
\fmfi{fermion,lab=$p$,lab.sid=right}{dd--gg}
\fmfi{dashes_arrow,lab=$\omega\tau$,lab.sid=left}{bb{down} .. tension 
5 .. {up}ii}
\fmfi{dashes_arrow,lab=$\omega\tau_{2}$,lab.sid=left}{cc{down} .. 
tension 3 .. {up}bb}
\fmfi{dashes_arrow,lab=$\omega\tau_{1}$,lab.sid=right}{dd{up} .. tension
3.. {down}cc}
\fmfi{dashes_arrow,lab=$\omega\tau'$,lab.sid=left}{ff{down} .. 
tension 5 .. {up}dd}
\end{fmfgraph*}
\vspace*{2cm}
\begin{center}
\caption{\label{f1}
Light-front time ordered graph for $ e^+e^- \rightarrow \gamma $. A
similar diagram with the opposite time-ordering for the photon exchange
should be added.}
\end{center}
\end{figure}

The off-energy-shell amplitude depicted in fig.\ref{f1}
  is given by the rules  of the
graph techniques \cite{cdkm} and  writes:
\begin{eqnarray}\label{eq1}
\bar{u}(p)M_1^{\rho}v(p')&=&e^2\int\bar{u}(p)\gamma_{\mu}(\hat{k}+m)
\theta(\omega\cd k)\delta(k^2-m^2)\frac{d^4k}{(2\pi)^3}
\\
&\times&  \gamma^{\rho}
\left(m-(\hat{Q}-\hat{k})\right)\theta(\omega\cd (Q-k))
\delta\left((Q-k+\omega\tau_2)^2-m^2\right)
\frac{d\tau_2}{\tau_2-i0} \nonumber\\ &\times&
\gamma_{\nu} v(p')(-g^{\mu\nu})
\delta\left((p-\omega\tau'+\omega\tau_1-k)^2-\mu^2\right)
\theta(\omega\cd (p-k))\frac{d\tau_1}{\tau_1-i0}.
\nonumber
\end{eqnarray}
where $Q=q-\omega \tau$. Note that
the fermion and antifermion propagators in LFD
differ from each other. The propagator $(\hat{k}+m)$ in (\ref{eq1})
corresponds to the electron, whereas the propagator
$\left(m-(\hat{Q}-\hat{k})\right)$ corresponds to the positron.

The factor $m-(\hat{Q}-\hat{k})=m-(\hat{k}_1-\hat{\omega}\tau_2)$
incorporates the difference $\hat{k}_1-\hat{\omega}\tau_2$ and,
therefore, takes into account the contact term $-\hat{\omega}\tau_2$.
We consider in this section the case $Q^2=4m^2$ relevant for the decay
width of the positronium, so that $p'=p$, $\tau'=0$ and so the two
graphes corresponding to the two different time orderings should give
the same contribution $M^{\rho}=M_1^{\rho} +M_2^{\rho}=2M_1^{\rho}$.

After integration over $\tau_{1}$ and  $\tau_{2}$, the amplitude
(\ref{eq1}) is given by:
\begin{eqnarray}\label{eq6}
\bar{u}(p)M_1^{\rho}v(p)&=&- e^2\int\bar{u}(p)\gamma_{\mu}(\hat{k}+m)
\gamma^{\rho}
\left(m-(\hat{Q}-\hat{k})\right)
\gamma^{\mu} v(p)
\nonumber\\
&\times&
\frac{\theta(\omega\cd (p-k))
\theta(\omega\cd (Q-k))
\theta(\omega\cd k)}
{(s_{12}-Q^2)(1-\frac{\omega\cd k}{\omega\cd Q})\;
(s_{123}-Q^2)
\left(\frac{\omega\cd p-\omega\cd k}{\omega\cd Q}\right)}
\frac{d^3k}{2\varepsilon_k(2\pi)^3}\ ,
\end{eqnarray}
where
$$
s_{12}-Q^2=2(\omega\cd Q)\tau_{2},\quad s_{123}-Q^2=2(\omega\cd
Q)\tau_{1}\ ,
$$ and
\begin{eqnarray}\label{eq7}
s_{12}&=&(k+k_1)^2=\frac{\vec{R}_{k\perp}^2+m^2}{x_{k}}
+\frac{\vec{R}_{k\perp}^2+m^2}{1-x_{k}}\ ,
\nonumber\\
s_{123}&=&(k+k_2+p)^2
=\frac{\vec{R}_{k\perp}^2+m^2}{x_k}
+\frac{\vec{R}_{k_1\perp}^2+\mu^2}{x_{k_1}}
+\frac{\vec{R}_{p'\perp}^2+m^2}{x_{p'}}
\nonumber\\
&=&\frac{\vec{R}_{k\perp}^2+m^2}{x_{k}}
+\frac{\vec{R}_{k\perp}^2+\mu^2}{1/2-x_{k}}+2m^2\ .
\end{eqnarray}
Like in the previous section, we define above the variables
$R_l=l-x_l Q$ with $x_l=\omega\cd l/\omega\cd Q$, where
$l$ is either $k, k_{1}$ or $p'$. At the threshold $Q^2=4m^2$,
  we have $\vec{R}_{p'\perp}=0$ and $x_{p'}=1/2$ in the variable $s_{123}$.  We
thus find:
\begin{equation}\label{eq8}
\tilde{M}^{\rho}=-2\frac{ e^2}{(2\pi)^3}\int d^2R_{\perp}
\int_0^{1/2}
\frac{O^{\rho}}
{(s_{12}-4m^2)(1-x)\;
(s_{123}-4m^2)\left(1/2-x\right)}\frac{dx}{2x}\ ,
\end{equation}
where:
\begin{equation}\label{eq9}
O^{\rho}=(\hat{p}+m)\gamma_{\mu}(\hat{k}+m)
\gamma^{\rho}\left(m-(\hat{Q}-\hat{k})\right)\gamma^{\mu}(\hat{p}-m)\ .
\end{equation}
The amplitude $\tilde{M}^{\rho}$ is connected to $M^{\rho}$ by eq.(\ref{eq3}).
The factor 2 in (\ref{eq8}) results from the sum of the two amplitudes
$M_1$ and $M_2$.
In order to find the coefficients $A,B,C$ which determine the amplitude
(\ref{eq2}), we
  substitute $\tilde{M}^{\rho}$ into eqs.(\ref{eq4}), (\ref{eq5}),
  regularize the expression by the Pauli-Villars prescription:
$$
A\to A(\mu) - A(\Lambda)\ ,
$$
(and similarly for $B,C$), and take the limits $\mu\to 0$,
$\Lambda\to\infty$.  The details of the calculation are given in the
appendix \ref{ap1}.  The final expression for $A$ is then:
\begin{equation}\label{eq18c}
A=\frac{\alpha m}{\mu}-\frac{7\alpha}{8\pi}
+\frac{\alpha}{2\pi}\log\left(\frac{\mu^2}{m^2}\right)
+\frac{\alpha}{4\pi}\log\left(\frac{\Lambda^2}{m^2}\right).
\end{equation}
This expression exactly coincides with that calculated in the Feynman
formalism  \cite{correc}.

The integral for $B(\mu)$ converges and does not depend on $\mu$:
\begin{equation}\label{eq21}
B(\mu)=-\frac{\alpha}{4\pi}\ .
\end{equation}
The amplitude regularized  \`a la Pauli-Villars is determined by the difference
$B_{reg}=B(\mu)-B(\Lambda)$. It is therefore zero.

  From (\ref{eq18c}) and (\ref{eq30a}) we recover  the
renormalized amplitude
\begin{equation}\label{eq32}
A_{ren}=A-Z=\frac{\alpha m}{\mu}-\frac{2\alpha}{\pi}\ ,
\end{equation}
which coincides with the result found in the Feynman approach \cite{correc}.
It contains the term $\frac{\alpha m}{\mu}$, corresponding to an infrared
singularity. In the next section we will show that this singularity
is physical but does
not contribute to the relativistic correction to the decay width of the
positronium.

The calculation of $C$ gives a divergent result even after a single 
Pauli-Villars
regularization. However, $C$ is the coefficient in front of the term
$\omega^{\rho}\bar{u}\hat{\omega}u$ which is proportional to
$\omega^\rho$. Like $Z'$ in eq. (\ref{eq23}),
it is separated out in eq. (\ref{eq2})  and does not contribute to the
observable amplitude, determined by $A$.

%%%%%%%%%%%%%%%%%%%%%%%%%%%%%%%%%%%%%%%%%%%%%%%%%%%%%%%%%%%%%%%%%%%%%%%%%%%%
\subsection{Infrared singularity}\label{infra}
The infrared singularity in (\ref{eq32}) results from the Coulomb
interaction between the electron and the positron.  It manifests
itself in the low momentum limit in the loop of
fig.2.

To calculate this limit, it is sufficient to take the nonrelativistic
limit for all the four momenta.
It means that we make in the numerator of
eq.(\ref{eq6}) the following replacements:
$$
\bar{u}(p)\gamma_{\mu}(\hat{k}+m) \to\bar{u}(p) (1+\gamma_0)m
\quad \mbox{and}\quad
\left(m-(\hat{Q}-\hat{k})\right)\gamma^{\mu}v(p)
\to
-m(1-\gamma_0)v(p).
$$
We used the fact that in this equation only the matrix $\gamma_{\mu}$ with
$\mu=0$
contributes. After this substitution,
  the integral (\ref{eq6}) converges
and its calculation, in the  $\mu\to 0$ limit,  gives
$$
2\bar{u}(p)M_1^{\rho}v(p)=\frac{\alpha m}{\mu}\bar{u}(p)\gamma^{\rho}v(p).
$$

This Coulomb contribution has to be removed from the total radiative
corrections (\ref{eq32}) since it is already  taken into account in 
the calculation of
the positronium wave function.
After this, we reproduce the well known radiative correction
$-\frac{2 \alpha}{\pi}$.

This result enables us to calculate the relativistic
radiative corrections beyond the Weisskopf-Van Royen approximation.
This is done for instance in ref.\cite{bdm} for the calculation of the
leptonic decay width of the $J/\psi$.

%%%%%%%%%%%%%%%%%%%%%%%%%%%%%%%%%%%%%%%%%%%%%%%%%%%%%%%%%%%%%%%%%%%%%%%%%%
\section{The electron self-energy}\label{sen}
Since our formulation of LFD is explicitly covariant, we are able to
follow very closely the standard procedure of renormalization of the
fermion self-energy in perturbation theory.  The self energy diagram
is shown in Fig.3.  As already done for the electromagnetic vertex, we
can write down immediately the general spin structure of the self
energy.  It is very simple and given by:
\begin{equation}
\label{eq2sen}
\Sigma(p) =A_{1}(p^{2})+B_{1}(p^{2})\frac{\hat{p}}{m}
+C_{1}(p^{2})\hat{\omega}\ .
\end{equation}
The coefficients \( A_{1},B_{1},C_{1} \) are scalar
functions of \( p^{2} \) only.
Here $p=p_{1}-\omega \tau _{1} $ is the total momentum entering
the diagram, $p_{1} $ is the external fermion momentum, with \(
p_{1}^{2}=m^{2} \), and \( \omega \tau _{1} \) is the external spurion
momentum. The off-energy-shell mass operator $\Sigma(p)$ depends on the
value  \( p^{2}=m^{2}-2(\omega \cd p)\tau _{1}.  \)

Like in the above calculations, we subtract from $\Sigma(p)$ the
$\omega$-dependent structure,
introducing:
\begin{equation}
\label{eq2sena}
\tilde \Sigma(p) =\Sigma(p)-C_{1}(p^{2})\hat{\omega}=
A_{1}(p^{2})+B_{1}(p^{2})\frac{\hat{p}}{m}\ .
\end{equation}
The standard procedure of renormalization of Feynman diagrams relies
on two counterterms: the mass counterterm $\delta m^2$ and the wave
function renormalization proportional to $Z_{2}$ \cite{IZ}.  They can
be incorporated explicitly in the Hamiltonian.

Following \cite{ll},
the renormalized self energy $\Sigma _{R}(p)$ is defined as
the part of $\tilde \Sigma(p)$ which is of second order  in the variable
$(\hat{p}-m)$. Without loss of generality, we can rewrite $\Sigma(p)$ 
in the form:
\begin{equation}
\label{eq7asen}
\tilde \Sigma(p)=A_{0}+(\hat{p}-m)B_{0}+\Sigma _{R}(p)\ .
\end{equation}
Here \( A_{0},B_{0} \) are constants
(they do not depend on \( p^{2} \))\ ,
and  $\Sigma _{R}(p)$ is the renormalized self-energy written as:
\begin{equation}\label{eq7ad}
\Sigma _{R}(p)=(\hat{p}-m)^{2}{\cal M}(p)\ ,
\end{equation}
where the matrix ${\cal M}(p)$ can be represented as:
\begin{equation}
\label{eq12sen}
{\cal M}(p)=a+(\hat{p}+m)b\ .
\end{equation}

\begin{figure}[hbtp]

\vspace*{2cm}
\hspace*{4 cm}
\begin{fmfgraph*}(100,40)
\fmfipair{aa,bb,cc,dd,ee,ff}
\fmfiequ{aa}{0.5[sw,nw]}
\fmfiequ{bb}{aa+(.5w,0)}
\fmfiequ{cc}{bb+(1.w,0)}
\fmfiequ{dd}{cc+(.5w,0)}
\fmfiequ{ee}{aa+(0,-.5h)}
\fmfiequ{ff}{dd+(0,-.5h)}
\fmfi{photon}{bb{up} .. tension 2 .. {down}cc}
\fmfi{fermion,lab=$p_{1}$,lab.sid=left}{aa--bb}
\fmfi{fermion,lab=$k$,lab.sid=left}{bb--cc}
\fmfi{fermion,lab=$p_{1}$,lab.sid=left}{cc--dd}
\fmfi{dashes_arrow,lab=$\omega\tau$,lab.sid=left}{cc{down} .. tension 
2 .. {up}bb}
\fmfi{dashes_arrow,lab=$\omega\tau_{1}$,lab.sid=left}{bb{down} .. tension
5 .. ee}
\fmfi{dashes_arrow,lab=$\omega\tau_{1}$,lab.sid=left}{ff .. tension
5 .. {up}cc}
\end{fmfgraph*}
\vspace*{1cm}
\begin{center}
\caption{\label{fig3}The electron self energy graph}
\end{center}
\end{figure}
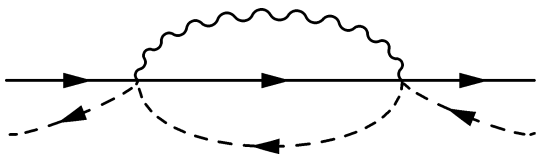
The renormalized fermion propagator now reads:
$$
\frac{1}
{\hat{p}-m-(\hat{p}-m)^{2}{\cal M}(p) }\ .
$$
It has the same pole and the same residue at $\hat p=m$ as the free propagator
of the physical fermion.

The explicit calculation of the renormalized fermion self energy
is now straightforward. According to the rules of CLFD, the electron
self energy shown on fig. \ref{fig3}  has the form:
\begin{eqnarray}
-\Sigma(p)  & = &  e^{2}\int \theta (\omega \cdot k)\delta
(k^{2}-m^{2})\gamma
^{\mu }(\hat{k}-\hat{\omega }\tau +m)\gamma ^{\nu} (-g_{\mu \nu})\nonumber
\label{eq1sen} \\  &
   & \times \theta (\omega \cdot (p-k))
\delta ((p+\omega \tau -k)^{2}-\mu
^{2})\frac{d\tau }{\tau -i0}\frac{d^{4}k}{(2\pi )^{3}}
\nonumber \\  & = &
-\frac{ e^{2}}{(2\pi )^{3}}
\int \frac{4m -2\hat{k}+2\hat{\omega }\tau
}{s-p^{2}}\frac{d^{2}R_{\perp}dx}{2x(1-x)}\ .
\end{eqnarray}
As indicated in the previous section, we introduce the photon mass
$\mu$ for infrared regularization.  The term $\hat{\omega }\tau$ in
(\ref{eq1sen}) contributes to $C_{1}(p^{2})\hat{\omega }$ only and can
be omitted in the calculation of $\tilde \Sigma(p)$.  In eq.(\ref{eq1sen}),
$\tau =(s-p^{2})/(2\omega \cd p)$ and
$$
s=\frac{R^{2}_{\perp}+m^{2}}{x}+\frac{R^{2}_{\perp}+\mu ^{2}}{1-x},\quad
k\cd p =\frac{R^{2}_{\perp}+m^{2}}{x}-\frac{1}{2}x p^2\ ,
$$
with $x=\omega\cd k/\omega\cdot p$, $R=k-x p$, and the phase-space volume is
given by
$d^3k/\varepsilon_k=d^2R_{\perp}dx/x$.

Now, starting from eq.(\ref{eq1sen}) we can calculate $\Sigma
_{R}(p)$, and the scalar coefficients $a$ and $b$ in
(\ref{eq12sen}).  Knowing $\Sigma(p)$ from eq.(\ref{eq1sen}), we
calculate the coefficients $A_1,B_1$ in (\ref{eq2sen}) and find
$\tilde \Sigma(p)$ by eq.(\ref{eq2sena}) .  Comparing (\ref{eq2sena})
with (\ref{eq7asen}), we express $A_0,B_0$ through $A_1,B_1$ for
$p^2=m^2$.  Using again the representation (\ref{eq7asen}) for
$\tilde \Sigma(p)$, we finally obtain the functions $a,b$, which
determine the self-energy (\ref{eq7ad}), through $A_1(p^2),B_1(p^2)$
and $A_0,B_0$.  The details of the calculation are given in appendix
\ref{ap2}.  For the functions $a$ and $b$ we find:
\begin{eqnarray}\label{eq16sen}
a & = & \frac{\alpha}{4\pi m}\frac{1}{(1-\rho)}
\left(1-\frac{2-3\rho}{1-\rho}\log\rho\right)\ ,
\nonumber\\
b & = &-\frac{\alpha}{2\pi m^2\rho}\left[\frac{1}{2(1-\rho)}
\left(2-\rho+\frac{\rho^2+4\rho-4}{1-\rho}\log\rho\right)
+1+\log\frac{\mu^2}{m^2}\right]\ ,
\end{eqnarray}
where
\[
\rho =\frac{m^{2} - p^{2}}{m^{2}}\ . \]
With these expressions for $a$ and $b$
the renormalized mass operator writes:
\begin{equation}\label{eq16b}
\Sigma _{R}(p)=(\hat{p}-m)^2{\cal M}(p)=
(\hat{p}-m)^2[a+(\hat{p}+m)b]\ .
\end{equation}
It exactly coincides with the standard result given for instance in
refs.\cite{IZ,ll}.  Note that in higher order calculations, the
$\omega$-dependent term $C_{1}(p^2) \hat \omega$ may give an $\omega$
independent contribution, when $\Sigma(p)$ enters as a part of a more
complex diagram.  It can also be renormalized.  The calculation of
$C_{1}^{ren}$ is given in appendix B. 

%%%%%%%%%%%%%%%%%%%%%%%%%%%%%%%%%%%%%%%%%%%%%%%%%%%%%%%%%%%%%
\subsection{The antifermion self-energy}
Since the forms of the fermion and antifermion propagators are
different in LFD (they contain $(\hat{p}+m)$ for fermion and $-(\hat{p}-m)$
for antifermion), the form of the self-energy is also different.
However, there is no need to repeat the calculation: the antifermion
self-energy $\overline{\Sigma}(p)$ can be found from the fermion one.

We represent $\overline{\Sigma}(p)$ similarly to (\ref{eq2sen}):
\begin{eqnarray}
\label{eq2a}
\overline{\Sigma}(p) &=&\overline{A}_{1}(p^{2})
+\overline{B}_{1}(p^{2})\frac{\hat{p}}{m}
+\overline{C}_{1}(p^{2})\hat{\omega }
\nonumber\\
&=&\overline{A}_{0}
+\overline{B}_{0}(\hat{p}+m)
+\overline{C}_{1}(p^{2})\hat{\omega }+\overline{\Sigma}_R(p)\ ,
\end{eqnarray}
where $\overline{A}_{0},\overline{B}_{0}$ are constants.
The renormalized antifermion self-energy $\overline{\Sigma}_R(p)$
is represented in the form similar to  (\ref{eq16b}):
\begin{equation}
\label{eq14a}
\overline{\Sigma} _{R}(p)=(\hat{p}+m)^{2}[\bar{a}-(\hat{p}-m)\bar{b}].
\end{equation}
One can easily see that
\begin{equation}
\label{eq14b}
\bar{a}=a\quad \bar{b}=b\ ,
\end{equation}
where $a,b$ are given by eq.(\ref{eq16sen}).
Indeed,
from the comparison of the fermion and antifermion propagators we
see that the antifermion self energy $\overline{\Sigma}_R(p)$
can be obtained from the fermion one $\Sigma_R(p)$ by:
\begin{equation}\label{eq1aa}
\overline{\Sigma}_R(p,m)=-\Sigma_R(p,-m)=-(\hat{p}+m)^{2}[a(-m)+(\hat{ 
p}-m)b(-m)]\ .
\end{equation}
Comparing  (\ref{eq14a}) with  (\ref{eq1aa}) and taking into account that
$a(-m)=-a(m)$, $b(-m)=b(m)$ (see the explicit expressions (\ref{eq16sen}))
we reproduce eqs. (\ref{eq14a}) and (\ref{eq14b}).

%%%%%%%%%%%%%%%%%%%%%%%%%%%%%%%%%%%%%%%%%%%%%%%%%%%%%%%%%%%%
\section{Conclusion}\label{concl}

The understanding of perturbative renormalization in QED is an
unavoidable step before studying more subtle systems like QCD. While
this perturbative renormalization is now a text-book section for the
standard formulation of field theory using Feynman graph techniques,
it is not as well understood in Light-Front Quantization. The main
reason being the difficulty to exhibit the covariant structure
of the electromagnetic vertex and electron self-energy since
standard LFD breaks explicitly covariance.

We have shown in this study that the covariant formulation of LFD is a
powerful tool in order to make the link between LFD and Feynman
approaches.  The explicit covariance of our formulation enables us to
exhibit the relativistic structure of the electromagnetic vertex in
QED, as well as the electron self-energy.  We are thus in a position
to extract, after renormalization, the finite physical contribution
from the infinite amplitude.  To do that, we have to know the
dependence of the operators on the orientation, $\omega$, of the light
front.  This is trivial in CLFD. In the standard formulation of LFD,
this dependence can not always be disentangled from the physical part of the
amplitude.  This $\omega$-dependent contribution is
responsible for the non-locality of the counter terms needed to
renormalize the infinite amplitude in LFD \cite{brs73}-\cite{glaz92}.

The finite physical amplitude we found in our approach for the
electromagnetic vertex, the electron self-energy, and the $e^+e^-\to
\gamma$ amplitude agree thus with  the standard text-book results.

We emphasize that in order to reproduce these results, a covariant
regularization of divergences (Pauli-Villars in the present study)
is important. The attempt to regularize the integrals by a cut-off in
the variable $R_{\perp}$,for instance, allows of course to work with finite
integrals.
It gives finite, but however wrong, renormalized results.

The QED vertex  $e^+e^-\to \gamma$, corrected by a color factor,
coincides with the QCD  vertex  $ q^+q^- \to \gamma$.
This result is applied in \cite{bdm} for the
calculation of the relativistic radiative correction to the $J/\psi$ leptonic
decay width.

The question of non-perturbative
renormalization, for scalar particles, will be adressed in a
forthcoming publication \cite{karm01}.

\section*{Acknowledgements}
One of the author (V.A.K.) is sincerely grateful for the warm hospitality
of the Laboratoire de Physique Corpusculaire, Universit\'e Blaise Pascal in
Clermont-Ferrand, where this work was performed. This work was partially
supported by the grant No. 99-02-17263 of the Russian Fund for Basic
Researches.

%%%%%%%%%%%%%%%%%%%%%%%%%%%%%%%%%%%%%%%%%%%%%%%%%%%%%%%%%%%%%%%%%%%%%%%%%%%%%%%
\appendix

\section{Calculation of $A,B$ and $C$}\label{ap1}

As explained in sect.\ref{ampl}, we substitute $\tilde{M}^{\rho}$
into eqs.(\ref{eq4}), (\ref{eq5}) in order to find the coefficients
$A,B,C$ determining the amplitude (\ref{eq2}). We thus find:
\begin{equation}\label{eq10}
(A,B,C)=-\frac{8\pi\alpha}{(2\pi)^3}\int d^2R_{\perp}
\int_0^{1/2}
\frac{(a,b,c)}
{(s_{12}-4m^2)(1-x)\; (s_{123}-4m^2)(1/2-x)}\frac{dx}{2x}\ ,
\end{equation}
with:
\begin{eqnarray}
a&=&t_2-t_1-t_3=\frac{1}{16m^2}
\left\{Tr\left[O^{\rho}\hat{\omega}\right]\frac{p_{\rho}}{\omega\cd p}
-Tr\left[O^{\rho}\gamma_{\rho}\right]
-Tr\left[\tilde{M}^{\rho}\hat{\omega}\right]
\frac{\omega_{\rho}m^2}{(\omega\cd p)^2}\right\}
\nonumber\\
&=&-\frac{1}{x}\left[\vec{R}_{\perp}^2(1 - 2x) + m^2(1 + 4x^2)\right],
\nonumber\\
b&=&t_1-t_2+3t_3=
\frac{1}{x}\left[\vec{R}_{\perp}^2(1 - 4x)+ m^2(1 - 2x)^2(1 + 2x)\right],
\nonumber\\
c&=&-t_1+3t_2-3t_3=
-\frac{1}{4m^2x^2}\left[\vec{R}_{\perp}^4+ 2m^2\vec{R}_{\perp}^2(1 - 8x^2)
+m^4(1 - 4x^2)^2\right].
\label{eq11}
\end{eqnarray}
To calculate the traces (\ref{eq11}), we need the following scalar products:
\begin{equation}\label{eq14}
k\cd Q=2m^2+(1-x)(s_{12}-4m^2),\quad
k\cd p=k\cd Q/2,\quad
p\cd Q=Q^2/2=2m^2.
\end{equation}
Substituting (\ref{eq11}) into (\ref{eq10}), we find:
\begin{eqnarray}
A(\mu)&=&\frac{8\pi\alpha}{(2\pi)^3}\int d^2R_{\perp}\int_0^{1/2}
\frac{\left[\vec{R}_{\perp}^2(1 - 2x) + m^2(1 + 4x^2)\right]}
{\left[\vec{R}_{\perp}^2 + m^2(1 - 2x)^2\right]
\left[\vec{R}_{\perp}^2 + m^2(1 - 2x)^2+2\mu^2 x\right]}dx\ ,
\label{eq15}\\
B(\mu)&=&-\frac{8\pi\alpha}{(2\pi)^3}\int d^2R_{\perp}\int_0^{1/2}
\frac{\left[\vec{R}_{\perp}^2(1 - 4x)+
   m^2(1 - 2x)^2(1 + 2x)\right]}
{\left[\vec{R}_{\perp}^2 + m^2(1 - 2x)^2\right]
\left[\vec{R}_{\perp}^2 + m^2(1 - 2x)^2+2\mu^2 x\right]}dx\ ,
\label{eq16}\\
C(\mu)&=&\frac{4\pi\alpha}{(2\pi)^3}\int d^2R_{\perp}\int_0^{1/2}
\frac{\left[\vec{R}_{\perp}^4+ 2m^2\vec{R}_{\perp}^2(1 - 8x^2)
+m^4(1 - 4x^2)^2\right]}
{2m^2x\left[\vec{R}_{\perp}^2 + m^2(1 - 2x)^2\right]
\left[\vec{R}_{\perp}^2 + m^2(1 - 2x)^2+2\mu^2 x\right]}dx\ .
\nonumber\\
&&
\label{eq17}
\end{eqnarray}
The integral (\ref{eq17}) for $C$, which is the coefficient
in  front of the structure proportional to $\omega_\rho$,
  diverges quadratically at ${R}_{\perp}\to\infty$ and
logarithmically at $x=0$.
The integral (\ref{eq15}) for $A$
logarithmically diverges at ${R}_{\perp}\to\infty$.
The integral (\ref{eq16}) for $B$ at ${R}_{\perp}\to\infty$
has the asymptotic expression:
$$
B(\mu)\propto\int_0^{\infty}
\frac{dR_\perp}{R_\perp}\int_0^{1/2}(1-4x)dx\ .
$$
Since the integral over $x$ is zero, $B(\mu)$ is finite.

The regularization of $A(\mu)$ proceeds as follows. The integral over
$x$ in $A(\mu)$ can be done analytically. In the Pauli-Villars
regularization scheme,
we should take the difference $A(\mu)-A(\Lambda)$ and calculate the
convergent integral over $R_{\perp}$.
Equivalently, but technically easier, we
calculate $A(\mu,L)$ with the cutoff $L$ in the
variable $ R_{\perp}$, take the
difference $A(\mu,L)-A(\Lambda,L)$ and then take the limit $L\to\infty$.
The result is analytic, but lengthy.
In the limits $\mu\to 0$ and $\Lambda\to \infty$
we obtain eq.(\ref{eq18c}), which
coincides with the result calculated in the Feynman formalism  \cite{correc}.

%%%%%%%%%%%%%%%%%%%%%%%%%%%%%%%%%%%%%%%%%%%%%%%%%%%%%%%%%%%%%%%%%%%%%% 
%%%%%%%%%%%%%%%%%
\section{Calculation of $a$,$b$ and $C_{1}^{ren}$}\label{ap2}
According to eqs.(\ref{eq7ad}) and (\ref{eq12sen}),
the self-energy $\Sigma _{R}(p)$
is determined by the scalar functions $a$ and $b$. As we will see below,
$a$ and $b$ are determined by the coefficients $A_1(p^2)$ and $B_1(p^2)$ in the
general decomposition (\ref{eq2sen}) and by their combinations in the limit
$p^2\to m^2$.

 From (\ref{eq2sen}) we find
the coefficients $A_{1}(p^{2})$ and $B_{1}(p^{2})$:
\begin{equation}
\label{eq4sen}
A_{1}=\frac{1}{4}Tr[\Sigma(p) ],\quad B_{1}=
\frac{m}{\omega \cd p}Tr[\Sigma(p) \hat{\omega }]\ .
\end{equation}
Substituting here eq.(\ref{eq1sen}) for $\Sigma(p)$ we get:
\begin{eqnarray}
A_{1}(p^2) & = & \frac{\alpha m}{\pi ^{2}}\int
\frac{\pi dR^{2}_{\perp}dx}{R^{2}_{\perp}+(1-x)m^{2}+x[\mu
^{2}+(1-x)p^{2}]}\ ,
\label{eq5sen} \\
B_{1}(p^2) & = & -\frac{\alpha m}{2\pi ^{2}}
\int \frac{\pi dR^{2}_{\perp}xdx}{R^{2}_{\perp}+(1-x)m^{2}
+x[\mu ^{2}+(1-x)p^{2}]}\ .
\label{eq6sen}
\end{eqnarray}
These integrals diverge logarithmically.

Comparing (\ref{eq2sen}) with (\ref{eq7asen}) and taking into account 
(\ref{eq7ad}),
we find:
\begin{equation}
\label{eq7sen}
A_{1}(p^{2})+B_{1}(p^{2})\frac{\hat{p}}{m}
=A_{0}+(\hat{p}-m)B_{0}
+(\hat{p}-m)^{2}{\cal M}(p)\ .
\end{equation}
 From here we can express the constants $A_0$ and $B_0$ through
$A_{1}$ and $B_{1}$:
\begin{eqnarray}
\label{eq8sen}
A_{0}&=&\frac{1}{4m}Tr\left[ \left( A_{1}(p^{2})+B_{1}(p^{2})
\frac{\hat{p}}{m}\right) (\hat{p}+m)\right] _{p^{2}=m^{2}}
=A_{1}(m^{2})+B_{1}(m^{2})
\\
\label{eq9sen}
B_{0}&=&\frac{1}{4m(p^{2}-m^{2})}
Tr\left[ \left( A_{1}(p^{2})+B_{1}(p^{2})\frac{\hat{p}}{m}-A_{0}\right)
(\hat{p}+m)^{2}\right] _{p^{2}\to m^{2}}\ .
  \end{eqnarray}
We thus obtain:
\begin{eqnarray}
A_{0} & = & \frac{\alpha m}{2\pi ^{2}}\int \frac{(2-x)\pi dR^{2}_{\perp}dx}
{R^{2}_{\perp}+(1-x)^{2}m^{2}+x\mu ^{2}}\label{eq10sen} \ ,\\
B_{0} & = & -\frac{\alpha }{2\pi ^{2}}
\int \frac{x[R^{2}_{\perp}-(3-4x+x^{2})m^{2}+x\mu ^{2}]\pi
dR^{2}_{\perp}dx}{[R^{2}_{\perp}+(1-x)^{2}m^{2}+x\mu ^{2}]^{2}}\ .
\label{eq11sen}
\end{eqnarray}
These integrals also diverge logarithmically.

 From (\ref{eq7sen}), and taking into account eq.(\ref{eq12sen}) for
${\cal M}(p)$, we get:
$$
(\hat{p}-m)^{2}(a+(\hat{p}+m)b)=A_{1}(p^{2})+B_{1}(p^{2})\frac{\hat{p}}{m}
-A_{0}-(\hat{p}-m)B_{0}\ .
$$
This allows finding out  $a$ and $b$:
\begin{eqnarray}
a & = & \frac{1}{4p^{2}(p^{2}-m^{2})}Tr\left[ \left(
A_{1}(p^{2})+B_{1}(p^{2})\frac{\hat{p}}{m}-A_{0}-(\hat{p}-m)B_{0}\right)
(\hat{p}+m)\hat{p}\right]
\nonumber\\
& = & \frac{A_1(p^{2})+B_{1}(p^{2})-A_0}{p^2-m^2}\ ,
\nonumber \label{eq13sen} \\
b & = &\frac{1}{4p^{2}(p^{2}-m^{2})^{2}}Tr\left[ \left(
A_{1}(p^{2})+B_{1}(p^{2})\frac{\hat{p}}{m}-A_{0}-(\hat{p}-m)B_{0}\right)
(\hat{p}+m)^{2}\hat{p}\right] \nonumber \\
\nonumber \\
& = &\frac{2m(A_1(p^{2})-A_0)}{(p^2-m^2)^2}
+\frac{(p^2+m^2)B_1(p^{2})}{m(p^2-m^2)^2}
-\frac{B_0}{p^2-m^2}\ .
   \end{eqnarray}
   Substituting
here the above expressions for $A_{1},B_{1},A_{0}$ and $B_{0}$, we get:
\begin{equation} \label{eq14sen}
a=\frac{\alpha m}{2\pi }\int
\frac{x(2-3x+x^{2})dR^{2}_{\perp}dx}{[R^{2}_{\perp}
+m^{2}(1-x)^{2}][R^{2}_{\perp}+m^{2}(1-x)(1-(1-\rho )x)]}\ .
\end{equation}
\begin{equation}
\label{eq15sen}
b=-\frac{\alpha}{2\pi }\int \frac{x^{2}(1-x)[R^{2}_{\perp}
-m^{2}(3-4x+x^{2})]dR^{2}_{\perp}dx}{[R^{2}_{\perp}+m^{2}(1-x)^{2}+\mu^2 x]^2
[R^{2}_{\perp}+m^{2}(1-x)(1-(1-\rho )x)+\mu^2 x]}\ .
\end{equation}
  We omitted in $a$ the photon mass \( \mu  \), since that integral has no
infrared divergence, and introduced the notation:
$\rho =(m^{2} - p^{2})/m^{2}$.
Integrating over $R^2_{\perp}$ and $x$ and keeping in $b$ the leading
term in $\log(\mu^2/m^2)$ only, we obtain eqs.(\ref{eq16sen}).

One can similarly calculate the coefficient $C_1$ determining the
$\omega$-dependent part of $\Sigma(p)$. It is given by:
\begin{eqnarray}\label{np1}
C_1(p^2)&=&\frac{1}{4\omega\cd p}Tr\left[\Sigma(p)\left(\hat{p}
-\frac{p^2\hat{\omega}}{\omega\cd p}\right)\right]
\nonumber\\
&=&-\frac{\alpha}{4\pi^2 \omega \cd p}\int
\frac{[2R_\perp^2+m^2(2-3(1-\rho)x^2)]\pi dR_\perp^2 dx}
{[R_\perp^2+m^2(1-x)(1-(1-\rho)x)+\mu^2x]x}
\end{eqnarray}
It is quadratically divergent in the variable $R_\perp$ and is
logarithmically divergent at $x=0$, despite the finite photon mass
$\mu$.  Note that the standard Pauli-Villars regularization is not
enough to make it finite.

We renormalize this scalar function in the standard way \cite{IZ}:
$$ C_1^{ren}(p^2)=C_1(p^2)-C_1(m^2)
-(p^2-m^2)\left.\frac{C_1(p^2)}{dp^2}\right|_{p^2=m^2}.
$$
After that it becomes finite:
\begin{eqnarray}\label{np2}
C_1^{ren}(p^2)&=&-\frac{\alpha m^4\rho^2}{4\pi^2 (\omega \cd p)}
\int\frac{[R_\perp^2(2-5x)+m^2(2-5x+3x^2)-3\mu^2x^2]x(1-x)\pi dR_\perp^2
dx} {[R_\perp^2+m^2(1-x)(1-(1-\rho)x)+\mu^2x]
[R_\perp^2+m^2(1-x)^2+\mu^2x]^2}
\nonumber\\
&=&-\frac{\alpha m^2\rho}{4\pi (\omega\cd p)}
\left[1-\frac{2-\rho}{2(1-\rho)}\log\rho +\log\frac{\mu^2}{m^2}\right].
\end{eqnarray}
%%%%%%%%%%%%%%%%%%%%%%%%%%%%%%%%%%%%%%%%%%%%%%%%%%%%%%%%%%%%%%%%

\end{fmffile}
\end{document}